\documentclass{iucrjournals}
\usepackage{graphicx}
\usepackage{siunitx}
\usepackage{lineno}

\usepackage[version=3]{mhchem}
\nolinenumbers
\title{2D-RIXS: Resonant inelastic x-ray scattering microscopy with high energy and spatial resolutions}

\author[a]{Kohei Yamamoto\IUCrEmaillink{kohei.yamamoto@qst.go.jp}\IUCrOrcidlink{0000-0002-4270-6410}}%
\author[b,c]{Hakuto Suzuki\IUCrEmaillink{hakuto.suzuki@tohoku.ac.jp}\IUCrOrcidlink{0000-0003-2973-0579}}%
\author[a]{Jun Miyawaki\IUCrEmaillink{miyawaki.jun@qst.go.jp}\IUCrOrcidlink{0000-0002-0602-907X}}

\affil[a]{NanoTerasu Center, National Institutes for Quantum Science and Technology, 468-1, Aoba, Aramaki, Aoba-ku, Sendai, Miyagi, 980-8572, Japan}
\affil[b]{Frontier Research Institute for Interdisciplinary Sciences, Tohoku University, Sendai 980-8578, Japan}
\affil[c]{Institute of Multidisciplinary Research for Advanced Materials (IMRAM), Tohoku University, Sendai 980-8577, Japan}

\begin{document} 
\maketitle 

\begin{synopsis}
We have established resonant inelastic x-ray scattering microscopy with the spatial resolution of \SI{1.0}{\micro\meter} combined with ultrahigh energy resolution in the soft x-ray regime.
\end{synopsis}

\begin{abstract}
A two-dimensional resonant inelastic x-ray scattering (2D-RIXS) microscopy system has been developed at the beamline BL02U of NanoTerasu. The instrument combines a Wolter type-I mirror for spatial imaging
with a varied-line-spacing grating spectrometer, 
simultaneously achieving micrometer-scale spatial resolution and ultrahigh energy resolution in the soft x-ray regime. 
Test chart measurements confirm a vertical spatial resolution of \SI{1.0}{\micro\meter} near the field-of-view center, and the horizontal resolution determined by the incident beam footprint is \SI{0.8}{\micro\meter}. 
RIXS imaging capabilities have been demonstrated by the measurements of a patterned NanoTerasu logo and exfoliated \ce{NiPS3} nanoflakes, 
highlighting its efficiency in locating specific microscale regions within inhomogeneous samples. 
These results establish 2D-RIXS microscopy as a position-sensitive probe of elementary excitations in quantum materials and functional devices.
\end{abstract}


\section{Introduction}
Resonant inelastic x-ray scattering (RIXS) is a versatile photon-in/photon-out spectroscopic technique capable of probing collective elementary excitations in quantum materials ~\cite{Ament2011,Ishii.K_etal.J.-Phys.-Soc.-Jpn.2013,de_Groot2024,Mitrano.M_etal.Phys.-Rev.-X2024}.
The dispersion relations in these collective excitations allow a microscopic description of low-energy quantum phenomena involving spin, charge, orbital, and lattice degrees of freedom in solids \cite{Imada.M_etal.Rev.-Mod.-Phys.1998,Tokura.Y_etal.Science2000}. Aided by significant advances in energy resolution \cite{Strocov2010,Dvorak2016,Brookes2018,Zhou2022,Singh2021}, RIXS has been widely applied to strongly correlated electron systems, including high-temperature superconductors and cathode materials in batteries.

Despite significant improvements in energy resolution in recent years, the real-space dependence of the RIXS response remains largely unexplored. Symmetry-broken phenomena, such as charge and magnetic ordering, commonly involve the formation of domains on mesoscopic length scales to lower the total free energy of the macroscopic specimen. Emergent functionalities in spintronic devices arise from the micro- to nano-scale engineering of magnetic materials. Moreover, recent advances in the fabrication of van der Waals (vdW) nanodevices and heterostructures have increased the demand for position-sensitive probes of their electronic states.
In this context, spatially resolved RIXS is expected to provide valuable insights into the electronic and magnetic properties of such materials and functional devices. 
However, achieving this goal requires the integration of two fundamentally different optical functions: spectroscopy and imaging. Consequently, instruments have typically been forced to specialize in one of them, sacrificing high resolution in the other dimension. 
This challenge is especially acute for an inherently photon-hungry technique like RIXS. 
The stringent requirements for high energy resolution alone are demanding, 
making the additional integration of an imaging functionality a formidable challenge.

Conventional RIXS spectrometers achieve higher energy resolution by using a narrower beamline exit slit to select monochromatic incident x-rays, 
a process that inherently sacrifices photon flux. 
The spectrometer then collects the scattered photons from the sample 
as a whole---without spatial resolution---to yield a single one-dimensional RIXS spectrum.
Although a two-dimensional detector is widely used to record the intensity of the scattered x-rays dispersed by the spectrometer gratings, 
the signal along the non-dispersive direction, which simply reflects 
the divergence of the scattered x-rays, has to be 
integrated to obtain the total intensity.
    
A known approach for spatially resolved RIXS, the so-called $h\nu^2$ concept~\cite{Strocov2010b,Zhou.K_etal.J.-Synchrotron-Rad.2020}, simultaneously addresses the limitation of conventional RIXS spectrometers. 
Instead of a slit-defined monochromatic beam, it utilizes an energy-dispersed line focus from the beamline on the sample, making more efficient use of the incoming x-rays. 
Correspondingly, the spectrometer images the scattered x-rays one-dimensionally along the direction of the line focus, repurposing the otherwise unused detector axis to record the spatial origin of the photons, while the perpendicular axis remains dedicated to the energy dispersion. 
This scheme enables the efficient and simultaneous acquisition of both position-sensitive and spectroscopic information.

This concept has been implemented in different optical designs.
One demonstration used a transmission off-axis Fresnel zone plate, which served as both a dispersive and an imaging element for the scattered x-rays~\cite{Marschall2017}.
While this approach achieved high spatial resolution, 
its energy resolution was limited to 804~meV in the full-width at half maximum (FWHM) 
at 400~eV, which is insufficient to resolve low-energy excitations.
A more common design, which our spectrometer also follows, combines Wolter type-I mirrors with a VLS grating spectrometer. 
Implementations at the European XFEL \cite{agaker_1d_2024} and Advanced Light Source \cite{Chuang2020} have prioritized spatial resolution, resulting in modest energy resolution.
Our 2D-RIXS spectrometer, on the other hand, has been developed to simultaneously achieve the ultrahigh energy resolution and high spatial resolution. Although it shares a similar optical layout with these previous instruments, our system achieves this dual purpose by combining Wolter type-I mirrors with a high-performance varied-line-spacing (VLS) grating spectrometer~\cite{Miyawaki2022}.
The ultrahigh energy resolution of this system has been demonstrated in previous publications \cite{Miyawaki2025, Yamamoto2025}. In the present paper, we report on the spatially resolved RIXS capability of the 2D-RIXS spectrometer, achieved without compromising its ultrahigh energy resolution.

\section{Experimental setup}
\begin{figure}[ht]
    
    \begin{center}
    \includegraphics[width=0.7\textwidth]{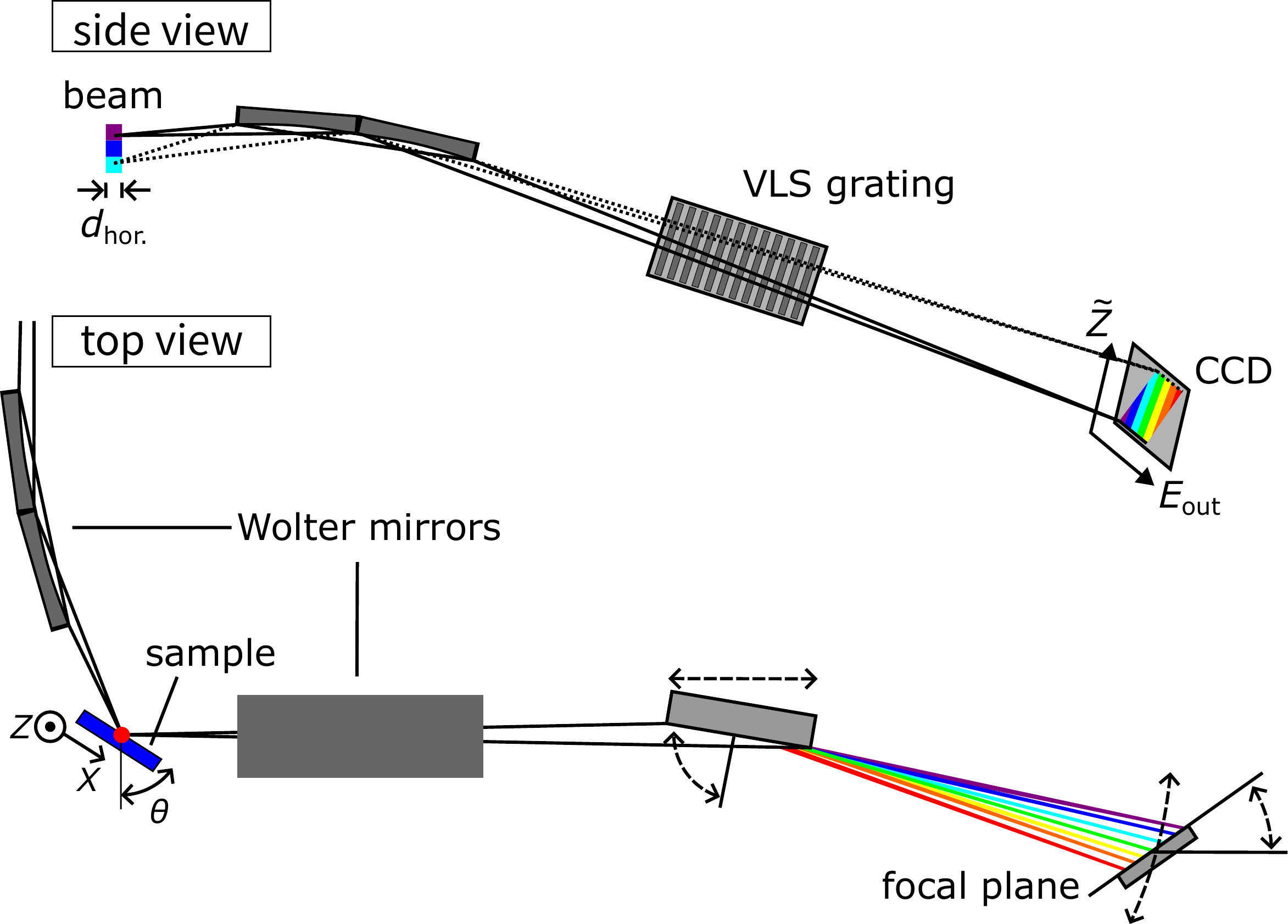}
    \end{center}
    \caption{
       Optical layout of the 2D-RIXS spectrometer.
       Dashed lines with bidirectional arrows indicate motor axes used for spectrometer alignment.}
       \label{fig:setup}
\end{figure}

The 2D-RIXS system has been developed and installed at the soft x-ray beamline BL02U of NanoTerasu.
Figure~\ref{fig:setup} shows the optical layout of the 2D-RIXS spectrometer.
In BL02U, x-rays from the undulator source pass through a focusing VLS plane grating monochromator (PGM, not shown). 
The PGM disperses these x-rays vertically according to their energy and focuses them vertically onto the sample position, 
which corresponds to the nominal exit slit position in a conventional beamline setup.
Simultaneously, a one-dimensional Wolter type-I mirror, placed after the PGM, 
focuses the x-rays horizontally onto the same sample position.
This optical arrangement delivers a vertically elongated, energy-dispersed line of focused x-rays onto the sample.
In Fig.~\ref{fig:setup}, this energy dispersion is schematically represented by the color gradient 
along the vertical direction at the sample position in the side view perspective, 
while the horizontal focus appears as a point in the top view.

The 2D-RIXS spectrometer analyzes the x-rays scattered from the sample by two main optical components arranged sequentially:
another one-dimensional Wolter type-I mirror providing vertical imaging, 
and a VLS grating providing horizontal energy dispersion (spectroscopy).
The Wolter mirror maps the vertically elongated line focus 
on the sample onto the 2D charge-coupled device (CCD) detector located \SI{12}{m} downstream, with a magnification ratio of 1:43. Hereafter, the vertical position in the CCD is represented by the scaled coordinate $\tilde{Z}$ (see the side view), whose value corresponds to the original distance at the sample position along the $Z$ axis (see the top view). 
As detailed in Sec.~\ref{sec:imaging_performance}, 
the imaging quality (spatial resolution) varies with the vertical position within the field of view (FOV).
Between the imaging Wolter mirror and the detector, 
the VLS grating disperses the scattered x-rays horizontally according to their energy 
and focuses them onto the two-dimensional detector.
The CCD detector (RIXScam, Xcam~Ltd.) is capable of single-photon detection in the soft x-ray regime and has a pixel size of \SI{16}{\micro\meter} and 
a total active area of $1608\times1650$ pixels (horizontal$\times$vertical).
By applying a centroid algorithm, 
the CCD can achieve sub-pixel spatial resolution of <\SI{5}{\micro\meter}.
Given the CCD dimensions and magnification, 
the geometric vertical FOV projected back onto the sample by the imaging Wolter mirror
is $\SI{16}{\micro\meter}\times1650/43=\SI{0.6}{mm}$.

To optimize performance for different experimental conditions (e.g. absorption edges), 
the spectrometer incorporates several mechanical adjustments. 
The grating can be translated along the optical axis over a range of approximately 1~m.
Additionally, the incident angle onto the grating can be changed by rotating the grating itself, 
and the included angle (the sum of the incident and diffraction angles at the grating) can be adjusted 
by moving the detector position laterally. 
Crucially, the control system ensures that the total distance from the sample to the detector 
remains fixed at 12~m to maintain the imaging condition, 
regardless of adjustments made to the grating position or the included angle.
Furthermore, the detector can be rotated around a vertical axis on its surface.
This adjustment allows the detector plane to match the focal curvature of the VLS grating, 
thereby optimizing the energy resolution across the detector.
Further details of the 2D-RIXS spectrometer are described elsewhere~\cite{Miyawaki2025}.

We first performed general beamline optical alignment, 
including the optimization of the horizontal focusing onto the sample position 
by fine-tuning the alignment parameters (e.g., position and incident angle) of the beamline's Wolter mirror. 
This ensured a sub-micrometer horizontal spot size of the incident x-rays on the sample surface.
Next, we optimized the total energy resolution of the combined system (beamline and spectrometer) by minimizing the line width of elastic scattering from a multilayer film placed at the sample position. 
We adjusted three key parameters iteratively:
(i) the $C_\textrm{ff}$ value (defined as $C_\textrm{ff} = \cos\beta / \cos\alpha$, 
where $\alpha$ and $\beta$ are the incidence and diffraction angles of the grating, respectively) of the beamline PGM grating,
(ii) the $C_\textrm{ff}$ value of the spectrometer VLS grating, and
(iii) the sample position along the beam direction.
Adjusting the sample position ensures that the sample surface is placed at the horizontal focus, 
which minimizes the beam footprint for microscopy 
and provides the smallest effective source size for the spectrometer required for optimizing the energy resolution.
Through this iterative optimization loop, 
we simultaneously achieved a total energy resolution of \SI{17.3}{meV} at the Ni $L_3$ edge (approx. 853 eV), 
and confirmed the horizontal spot size at the sample position to be $d_\textrm{hor.}=\SI{0.8}{\micro\meter}$ in FWHM.
While the vertical spatial resolution is primarily determined by the imaging Wolter mirror 
(discussed in Sec.~\ref{sec:imaging_performance}), 
the horizontal spatial resolution depends on the horizontal footprint size on the sample surface, 
which corresponds to the incident spot size ($d_\textrm{hor.}$) projected by the incident angle $\theta$, i.e., $d_\textrm{hor.}/\sin\theta$. 
%18.6*0.93=17.3meV
The optimal beamline conditions thus achieved allow RIXS microscopy experiments on different target materials as detailed below.

\section{Imaging mirror performance}\label{sec:imaging_performance}
\begin{figure}[ht]
    
    \begin{center}
    \includegraphics[width=0.9\textwidth]{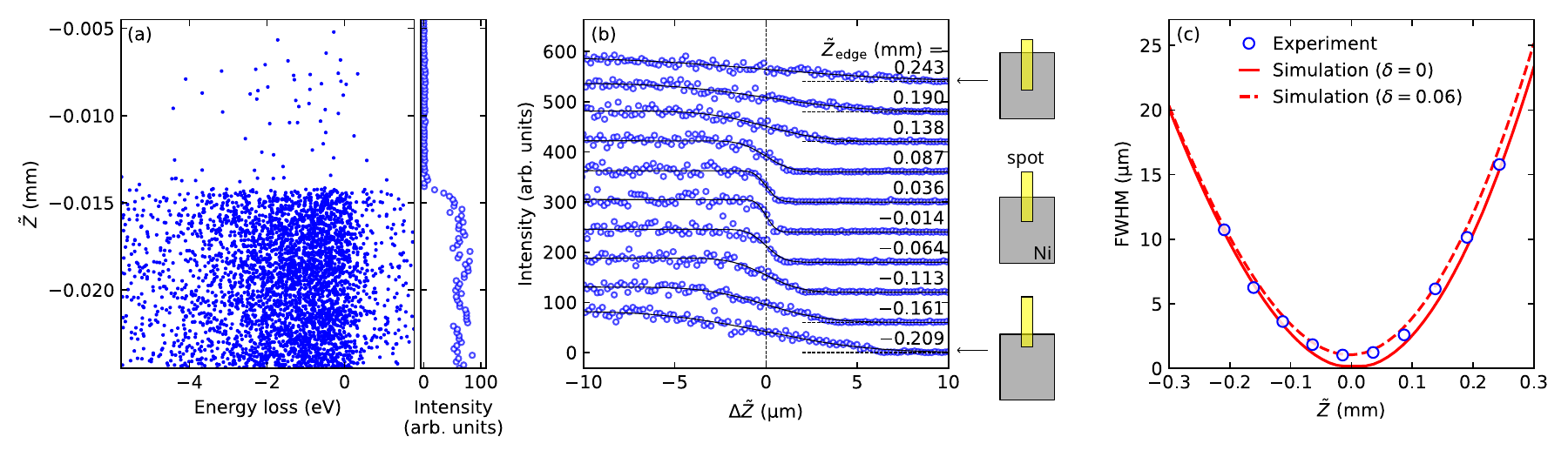}
    \end{center}
    \caption{
        (a) Scatter plot of detected photons around the sharp edge of the test chart made of Co thin film, as a function of photon energy and vertical position $\tilde{Z}$ around $\tilde{Z}=\SI{-0.014}{mm}$.
        Right panel shows the corresponding line profile.
        (b) Line profiles measured at different $Z$ positions. 
        The solid lines are the fitted curves by an error function. 
        $\tilde{Z}_\textrm{edge}$ indicates the center of the fitted error function and the line shape is plotted as a function of deviation from $\tilde{Z}_\textrm{edge}$, $\Delta \tilde{Z} = \tilde{Z}-\tilde{Z}_\textrm{edge}$.
        $\tilde{Z}_\textrm{edge}$ was changed by moving the test pattern.
        (b) The estimated resolution as a function of $\tilde{Z}$ position.
        The solid line shows the resolution determined by ray-tracing simulation (offset from the ideal Wolter mirror position $\delta=0$), while the dashed line shows the resolution considering the finite alignment error ($\delta=\SI{0.06}{mm}$).
    }\label{fig:z_edge}
\end{figure}

We first characterize the system performance by evaluating the intrinsic vertical spatial resolution, 
which is primarily determined by the imaging Wolter mirror located after the sample. 
This assessment utilized the standard technique of measuring the profile across a sharp edge of a test pattern.
An edge provides a well-defined step function in signal intensity in the CCD image (see Fig. \ref{fig:z_edge}(a)), 
which allows the determination of the spatial resolution from the sharpness of the integrated line profile (right panel).

The test chart was fabricated using a photolithography on a MgO(001) wafer
and forms a square pattern ($\SI{1.0}{mm}\times\SI{1.5}{mm}$, horizontal $\times$ vertical) 
made of a \SI{30}{nm}-thick Co metal film capped with a \SI{2}{nm}-thick \ce{SiO2} layer. 
The incident x-ray energy was tuned to the Co~$L_3$ absorption edge. 
We set the incident angle of the x-rays onto the sample surface to $\theta = \ang{65}$ and the scattering angle to $2\theta = \ang{90}$. 
In this geometry, the projected horizontal spatial resolution is given by $d_\textrm{hor.}/\sin\theta=\SI{0.9}{\micro\meter}$ in FWHM.
To evaluate the resolution across the FOV, we positioned the sharp horizontal edge of the square-shaped pattern 
at ten different vertical positions by moving the test chart along the $Z$ axis, as schematically shown in the right part of Fig.~\ref{fig:z_edge}(b). 
At each position, we recorded the image of the edge profile formed by the Wolter mirror onto the CCD. 
The line profile was then obtained by integrating the intensity along the energy-dispersive (horizontal) axis of the detector.

Figure~\ref{fig:z_edge}(b) displays the resulting line profiles, collected at different $Z$ positions of the chart. 
A clear edge step was observed in the entire range, with different sharpness.
To quantify the spatial resolution, we assumed a Gaussian line spread function (LSF) with a standard deviation of $\sigma$ and 
each lineshape was fitted with the 
edge spread function (ESF)
\begin{align*}
    \mathrm{ESF}(\tilde{Z}) &= \int^\infty_{-\infty} \mathrm{LSF}(\tilde{Z}-\tilde{Z}')\cdot A\cdot\mathrm{H}(\tilde{Z}'-\tilde{Z}_\mathrm{edge})d\tilde{Z}' %\\
    %&=\int_{\tilde{Z}_\mathrm{edge}}A\cdot\mathrm{LSF}(\tilde{Z}-\tilde{Z}')dz' \\
    = A\left[\frac{1}{2}+\mathrm{erf}\left(\frac{\tilde{Z}-\tilde{Z}_\mathrm{edge}}{\sigma}\right)\right],
\end{align*}

where  $\mathrm{H}$ is the Heaviside step function, $\mathrm{erf}$ is the error function, and $A$ is the intensity.
The spatial resolution is given by $2\sqrt{2\log 2}\cdot\sigma$, which corresponds to the FWHM of the Gaussian broadening.
The vertical edge position $\tilde{Z}_\textrm{edge}$ was determined as the center of the ESF, and the intensity profile is plotted as a function of the positional deviation from the center, $\Delta \tilde{Z} = \tilde{Z}-\tilde{Z}_\textrm{edge}$ in Fig.~\ref{fig:z_edge}(b).

The obtained vertical spatial resolution is plotted as a function of $\tilde{Z}$ in Fig.~\ref{fig:z_edge}(c). 
The best resolution achieved was \SI{1.0}{\micro\meter} near the center of the FOV ($\tilde{Z} \approx 0$). 
Meanwhile, the spatial resolution degrades significantly towards the edges, 
reaching approximately \SI{20}{\micro\meter} at $\tilde{Z} = \pm \SI{0.3}{mm}$. 
This broadening in the off-axis region is a characteristic feature of the geometric aberrations inherent in the Wolter mirror optics.

To gain a deeper understanding of the observed imaging performance and the factors limiting the resolution, 
we compared the experimental results with detailed ray-tracing simulations of the Wolter mirror optics. 
The solid line in Fig.~\ref{fig:z_edge}(c) illustrates the simulated resolution profile 
assuming perfect alignment of the Wolter mirror relative to the sample and detector, 
while incorporating the mirror's specified slope error and assuming an intrinsic detector spatial resolution of \SI{4}{\micro\meter}. 
This ideal alignment scenario predicts a 
resolution of \SI{0.165}{\micro\meter} at the center ($\tilde{Z}=0$).  
The discrepancy between the simulation and the experimentally achieved value of \SI{1.0}{\micro\meter} 
can be explained by a slight misalignment of the imaging mirror in the current instrument setup. 

Indeed, further ray-tracing simulations incorporating a small, physically plausible misalignment 
provide excellent agreement with the experimental data. 
As shown by the dashed line in Fig.~\ref{fig:z_edge}(c), introducing a small longitudinal offset of the mirror along the optical axis 
($\delta=\SI{0.06}{mm}$, corresponding to the mirror being positioned \SI{0.06}{mm} further away from the sample than ideal) 
accurately reproduces the experimentally observed resolution profile, 
including the spatial resolution degradation near the edges. 
This misalignment likely stems from the design decision: 
the adjustment mechanism required to correct this offset was intentionally omitted 
to prioritize mechanical stability for ultrahigh energy resolution. 
Importantly, this slight mirror misalignment, while limiting the vertical spatial resolution to the micrometer scale, 
does not compromise the spectrometer's performance in the energy dispersion direction. 
The vertical spatial resolution required to maintain the current ultrahigh energy resolution setup 
is estimated from simulations to be around \SIrange{4}{5}{\micro\meter}. 
The achieved resolution of \SI{1.0}{\micro\meter} near the center is well within this requirement. 
Even considering the vertical range illuminated by the energy-dispersed incident beam (approximately $\pm \SI{100}{\micro\meter}$ around the center), 
the spatial resolution remains better than \SI{5}{\micro\meter} (see Fig.~\ref{fig:z_edge}(b)). 
Therefore, as intended by the design, the ultrahigh energy resolution remains unaffected by the current imaging performance limitation.

\section{RIXS imaging of a patterned logo} \label{sec:RIXS_imaging}
\begin{figure}[ht]
    
    \begin{center}
    \includegraphics[width=0.7\textwidth]{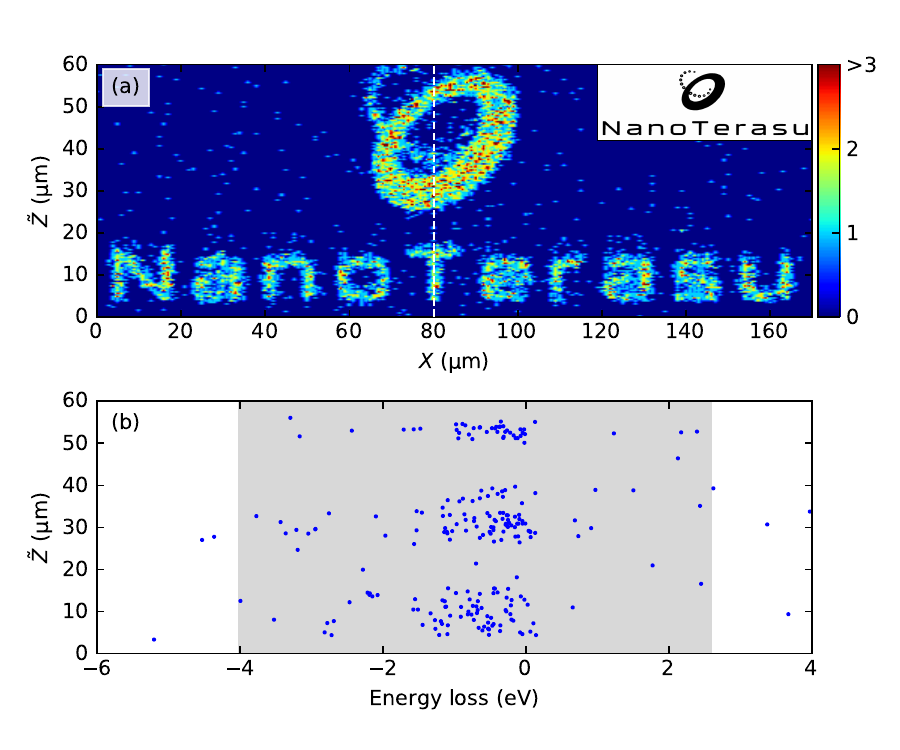}
    \end{center}
    \caption{
        (a) RIXS image of the patterned NanoTerasu logo collected at the Ni $L_3$ edge. The dashed line indicates horizontal position ($X=\SI{80}{\micro\meter}$) where the scatter plot (b) was obtained.
        Inset: Design of the Ni pattern fabricated on the substrate. 
        (b) Scatter plot of detected x-ray photons at $X=\SI{80}{\micro\meter}$ as a function of photon energy and vertical position $\tilde{Z}$.
        The RIXS colormap in (a) was constructed by integrating the photon counts within the gray-shaded loss-energy window.
        }\label{fig:test_pattern}
\end{figure}

We now demonstrate the core capability of the instrument: RIXS imaging. 
For this demonstration, we used a test pattern in the shape of the NanoTerasu logo (the inset of ~\ref{fig:test_pattern}(a)), 
which incorporates fine structures relevant to microscale devices.
The test pattern 
was fabricated using photolithography on a \ce{MgO}(001) substrate. 
A \ce{SiO2}~(\SI{2}{nm})/\ce{Ni}~(\SI{30}{nm}) thin film was deposited onto the substrate to form the logo pattern.
For the RIXS measurements, the incident photon energy was tuned to the Ni~$L_3$ absorption edge. 
The incident and scattering angles were set to $\theta=\ang{65}$ and $2\theta=\ang{150}$, respectively. 

To acquire the RIXS image shown in Fig.~\ref{fig:test_pattern}(a), 
the sample was scanned horizontally along the $X$ direction 
(perpendicular to the incident beam's line focus), 
over a range of \SI{170}{\micro\meter} in steps of $\Delta X = \SI{1}{\micro\meter}$. 
At each $X$ position, we recorded the scattered x-ray intensity 
as a function of both photon energy (dispersed horizontally on the detector) and 
vertical position $\tilde{Z}$ (imaged vertically on the detector), with an exposure time of \SI{60}{\second} per step.
The raw data obtained at $X=\SI{80}{\micro\meter}$ (indicated by the dashed line in Fig.~\ref{fig:test_pattern}(a))
is shown in Fig.~\ref{fig:test_pattern}(b), which contains the intensity variations corresponding to the pattern features along the vertical ($\tilde{Z}$) axis, 
as well as the electronic excitations of Ni metal along the horizontal axis.

To construct the final RIXS image shown in Fig.~\ref{fig:test_pattern}(a), 
we processed the data collected at each $X$ position. 
Specifically, for each $X$ positions, a 1D intensity profile along $\tilde{Z}$ direction 
was obtained by integrating the spectral weight within an 
energy loss window 
covering both elastic and inelastic features, indicated by the gray-shaded region in Fig.~\ref{fig:test_pattern}(b). 
These 1D profiles from all $X$ positions were arranged to 
construct the 2D ($X, \tilde{Z}$) image.

The resulting RIXS image in Fig.~\ref{fig:test_pattern}(a) clearly reproduces the fine features of the designed NanoTerasu logo pattern. 
The pronounced contrast between the Ni pattern (bright regions) and the MgO substrate (dark regions) confirms that the system successfully provides spatially resolved RIXS spectra with a sufficient signal-to-noise ratio and spatial fidelity to map the elemental variations across the sample surface. This measurement validates the capability of our 2D-RIXS spectrometer to perform RIXS microscopy measurements.

\section{Application to exfoliated van der Waals nanosheets} \label{sec:NiPS3_application}
To demonstrate its practical application, we showcase the use of the 2D-RIXS system for measuring the exfoliated nanosheets of vdW magnets. The development of mechanical exfoliation techniques for fabricating atomically thin layers has greatly advanced research on two-dimensional vdW materials and functional devices constructed from their heterostructures. In recent years,  interest has expanded beyond their characteristic single-particle electronic states—such as the massless Dirac electrons in graphene—to the study of two-dimensional quantum magnetism in vdW magnets \cite{Burch.K_etal.Nature2018}, which are now being explored for spintronic applications \cite{Zhang.B_etal.npj-Spintronics2024}. 

\begin{figure}[ht]
    \begin{center}
    \includegraphics[width=.8\textwidth]{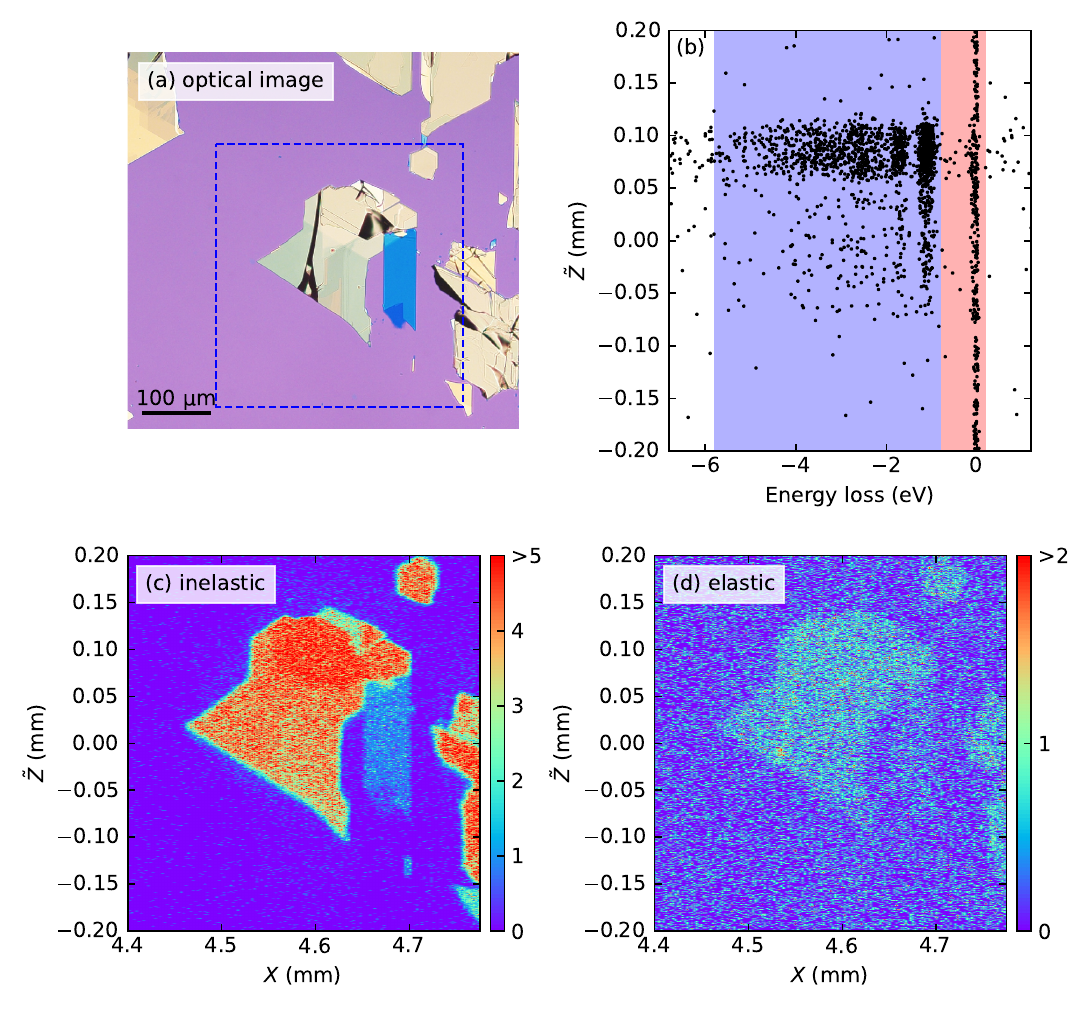}
    \end{center}
    \caption{
        (a) Optical microscope image of exfoliated \ce{NiPS3} nanoflakes.
        The dashed square indicates the region investigated by the RIXS imaging. 
        (b) Example scatter plot showing detected x-ray photons versus photon energy (horizontal axis) and vertical position $\tilde{Z}$ (vertical axis) 
        at a single horizontal position $X$.
        Blue- and red-shaded areas indicate the energy integration windows used to generate the maps in (c) and (d), respectively.
        (c, d) Colormap of RIXS intensity of the \ce{NiPS3} nanoflakes acquired at the Ni $L_3$ edge, 
        obtained by integrating photon counts within the (c) inelastic   
        and (d) elastic regions, respectively.
    } \label{fig:nips3}
\end{figure}

Since the scattering cross-section of RIXS is larger than those of other inelastic scattering techniques, 
such as inelastic neutron scattering, it is possible to detect signals from vdW nanosheets with tiny sample volumes 
\cite{Yang.Z_etal.Phys.-Rev.-B2023,Pelliciari.J_etal.Nat.-Mater.2024,Wei.Y_etal.npj-Quantum-Mater.2025}. 
However, the primary technical challenge for RIXS measurement is that the mechanical exfoliation often generates multiple flakes of different thicknesses randomly dispersed over the substrate. 
Figure~\ref{fig:nips3}(a) shows an optical microscope image of the \ce{NiPS3} nanosheets used in our study. The golden regions correspond to bulk-like flakes with large thickness, the blue regions to thinner ones, and, in particular, the dark-blue triangular area represents the thinnest region. If one aims to measure such ultrathin regions, a major technical challenge is that the target area occupies only a small fraction of the overall sample surface and is typically surrounded by much thicker flakes. 
As a result, it is highly nontrivial to differentiate signals from ultrathin and thick regions and to precisely irradiate x-rays onto the limited areas. 
In conventional RIXS spectrometers, a possible approach is to perform point-by-point 2D mapping of the sample surface to visualize the shape of the flakes \cite{Wei.Y_etal.npj-Quantum-Mater.2025}. However, this process is time-consuming and crucially constrained by the low spatial resolution determined by the x-ray spot size.

The imaging capability of the 2D-RIXS spectrometer provides an efficient alternative to locate the desired region.
Since the 2D-RIXS spectrometer inherently records the vertical position ($\tilde{Z}$) of scattered photons, 
performing a single scan along the horizontal ($X$) direction is sufficient to generate a 2D intensity map 
covering the region indicated by the dashed square in Fig.~\ref{fig:nips3}(a). 
During this scan, at each $X$ position, 
the scattered photons were recorded as a function of photon energy and vertical position $\tilde{Z}$, as shown in Fig.~\ref{fig:nips3}(b). The measurements were performed at the normal incidence $\theta=\ang{90}$ to minimize the beam footprint, and at the scattering angle of $2\theta = \ang{150}$.

By counting the photon events within specific energy-loss windows indicated in Fig.~\ref{fig:nips3}(b) for individual ($X, \tilde{Z}$) points, 
we constructed spatial RIXS intensity maps. 
Figures~\ref{fig:nips3}(c) and (d) display two intensity maps, 
generated by integrating the inelastic (\textit{d-d} excitations, blue-shaded region in Fig.~\ref{fig:nips3}(b)) and 
elastic (red-shaded) regions, respectively. 
The morphology revealed in both maps clearly corresponds to the flake distribution seen in the optical image. 
Furthermore, the intensity variations within the maps correspond systematically with the different flake thicknesses 
observed as the different colors in the optical image. 
Notably, the map generated from the strong \textit{d-d} excitations (Fig.~\ref{fig:nips3}(c)) exhibits a clearer contrast 
than that from the elastic scattering (Fig.~\ref{fig:nips3}(d)), as the former is not affected by the non-resonant elastic scattering from the substrate.  
This highlights the efficiency of the 2D-RIXS spectrometer in locating and spectroscopically characterizing 
specific microscale regions within inhomogeneous samples.

\section{Conclusion}
In summary, we have successfully developed and demonstrated the capabilities of a spatially resolved 2D-RIXS system at the beamline BL02U of NanoTerasu. 
The 2D-RIXS spectrometer combines a Wolter type-I mirror for vertical imaging with a high-performance VLS grating RIXS spectrometer for horizontal energy dispersion, enabling simultaneous achievement of high energy resolution and micrometer-scale spatial resolution in the soft x-ray regime.
Characterization using a test chart confirmed a best vertical spatial resolution of \SI{1.0}{\micro\meter} near the center of the FOV, 
while the horizontal resolution determined by the incident beam footprint was \SI{0.8}{\micro\meter}. 
We have demonstrated its imaging capabilities by RIXS intensity mapping of a patterned Ni film logo and 
exfoliated \ce{NiPS3} nanoflakes. 
These results establish 2D-RIXS microscopy, combining ultrahigh energy resolution with micrometer spatial sensitivity, 
as an efficient real-space probe of elementary excitations in quantum materials and functional devices.

\section{Acknowledgments}
We acknowledge the technical support from Takuya~Oki, Ralph~Ugalino and Kentaro~Fujii during the spectrometer commissioning.
We are grateful to Seiji~Sakai and Li~Songtian for providing the test chart, and Je-Geun~Park for providing the \ce{NiPS3} nanoflakes.
The RIXS experiments were performed using the 2D-RIXS spectrometer at the BL02U of NanoTerasu with the approval of the Japan Synchrotron Radiation Research Institute (JASRI) (Proposal No. 2025A9006, 2025A9059, 2025B9023, 2025B9047).

\begin{funding}
This work is supported by JSPS KAKENHI Grant Numbers JP22K13994 and JP25K00014.
\end{funding}

\ConflictsOfInterest{There is no conflict of interest.}

\DataAvailability{
The data supporting the results reported in this paper are available from the corresponding authors upon reasonable request.
}

\bibliography{RIXSimaging_JSR} % basename of .bib file

\end{document}